\documentclass[%
 reprint,
notitlepage,
 amsmath,amssymb,
 aps,
 onecolumn,
]{revtex4-1}

\usepackage{graphicx}
\usepackage{dcolumn}
\usepackage{bm}
\usepackage{hyperref}
\usepackage{times}
\usepackage{xcolor,import}
\usepackage{lipsum}

\usepackage[
margin=1in,
]{geometry}

\usepackage[T1]{fontenc}
\usepackage[utf8]{inputenc}
\usepackage{color}
\usepackage{url}
\usepackage{SIunits}
\usepackage{amssymb,amsmath,amsfonts}
\usepackage{subfigure}
\usepackage{xcolor}
\usepackage{xspace}

\usepackage{ulem}

\begin{document}

\title{Influence of surfactant concentration on drop production by bubble bursting}

\author{Juliette Pierre}
\email[]{juliette.pierre@sorbonne-universite.fr}
\author{Mathis Poujol}
\author{Thomas S\'eon}

\affiliation{%
Sorbonne Université, CNRS, UMR 7190, Institut Jean Le Rond $\partial$'Alembert,  F-75005 Paris, France
}

\date{\today}

\begin{abstract}

Bubble bursting at the surface of the sea water produce drops and is the main source of sea spay aerosol.
The mechanisms underlying the drops production from a single bubble bursting event have been intensively studied and the influence of the bubble size and liquid parameters (density, viscosity and surface tension) has been unified. However, despite the diversity of the surfactant molecules present in the oceans, their influence has been overlooked.
In this paper we experimentally explore the influence of the surfactant concentration (Sodium Dodecyl Surfate - SDS) in a water solution on a single bubble collapse and subsequent drop production.
We show that these surfactant molecules have an astonishing effect. In particular, we quantitatively show that they modify the bubble collapse, they induce less, smaller and faster drops and they can even completely prevent the drop production for a particular concentration.
These results are presented in the existing dimensionless framework and they allow us to affirme that these effects are mainly a consequence of the surface tension gradients (Marangoni stresses) more than just the surface tension lowering. 
Finally, this study shows that the role of water contamination by surface-active agents is important and needs to be understood to improve the prediction of the sea spray aerosol.

\end{abstract}


\pacs{Valid PACS appear here}

\maketitle



\section{Introduction}

As a wave crashes in the ocean it entrains air below the surface. After a turbulent break-up cascade \cite{Garrett2000} a population of bubbles is produced \cite{Deane2002, Prather2013, Deike2016} and while small bubbles may be dissolved into the water, larger bubbles rise back to the surface and collapse \cite{Deike2021}. The bursting of a bubble starts with the break up of the thin liquid film that separates the air cavity from the atmosphere and ends up with the fragmentation of a rising jet. Through these two fragmentation events, bubble bursting produces film drops \cite{Blanchard1988, Lhuissier2011b} and jet drops \cite{Spiel1997, Ghabache2014, Ghabache2016a, Brasz2018a, Ganan-Calvo2017, Blanco--Rodriguez2020} and constitutes one of the main sources of the ocean spray  \cite{Veron2012}. 
By evaporating, the sea spray transports in the atmosphere water vapor, important for the thermodynamics of the atmosphere, and salt crystals that affect the radiative balance of the atmosphere and form cloud condensation nuclei \cite{Lewis2004, Veron2015}. And, in a significant way, they carry also heat, dissolved gases, surfactants, biological materials \cite{Deike2021}.
Finally, uncertainties in predicting sea spray aerosols characteristics directly impacts our ability to perform weather prediction and earth system modeling \cite{Leeuw2011, Deike2018a}. 

Since the pioneer work of D. Blanchard \cite{Blanchard1963} there have been a number of - experimental, numerical and theoretical - combining studies on a single bubble bursting, that brought comprehensive data on the size and speed of the jet drops produced by bubble bursting in water \cite{Seon2017, Duchemin2002, Berny2020, Ganan-Calvo2018}. Applying these results to the bubble size distribution produced under a breaking wave enabled a rough estimation of the statistics of jet drop production \cite{Berny2021}.
However, the ocean surface is partly covered by a biofilm, which can be modeled with surfactants \cite{Wurl2011}. The surface-active contaminations are known to modify the static and dynamic behaviors of bubbles, including their coalescence, lifetimes, and bursting \cite{Poulain2018a, Shaw2021, Neel2021}. Consequently, the influence of the physicochemistry of the interface has to be taken into account in the process of bubble collapse at the interface and in the subsequent drop production. 
Néel \& Deike (2021) \cite{Neel2021} considered a monodisperse assembly of millimetric air bubbles produced identically in the bulk for a wide range of surface contamination and showed that, depending on the contamination, the bubble distribution that bursts can be very distinct from the initial distribution. 
There have been various experiments that attempted to described the role of the physicochemical parameters on the production of droplets by bursting bubbles \cite{Modini2013, Prather2013, Quinn2015, Deike2021}, but there are large variations in protocols, and the influence of surfactants on the drop production remains largely unclear. 
All these experiments are realized on a large collection of bubbles, with different size distributions, suggesting the need to carry out  a study on a single bursting bubble.
Recently, Constante-Amores {\it et al.} (2021) \cite{Constante-Amores2021} studied the effect of surfactant on the dynamics of a bubble bursting, using numerical simulations, accounting for sorption kinetics and diffusive effects. At one fixed bubble size and one surface contamination, they showed that the presence of surfactant affects the dynamics of the system through Marangoni-induced flow and is responsible for delaying the collapse and generating slower and fewer drops. 
 
In this article, we study experimentally the effect of Sodium Dodecyl Sulfate (SDS) surfactant on the dynamics of a bubble bursting through an interface. 
After describing the experimental setup, we show qualitatively that the surfactants have an astonishing influence on the jet dynamics subsequent to the bubble collapse, and on the jet drops production. The following is dedicated to quantify this effect by varying the surfactant concentration and the bubble size. We start by studying the influence of the surfactants on the bubble collapse time, before characterizing the variation of the number, size and speed of the ejected drops as a function of the control parameters. Finally, we focus on the influence of the surfactant concentration on the cavity collapse and the capillary waves dynamics.


\begin{figure}[b!]
    \centering
    \includegraphics[width=1.\linewidth]{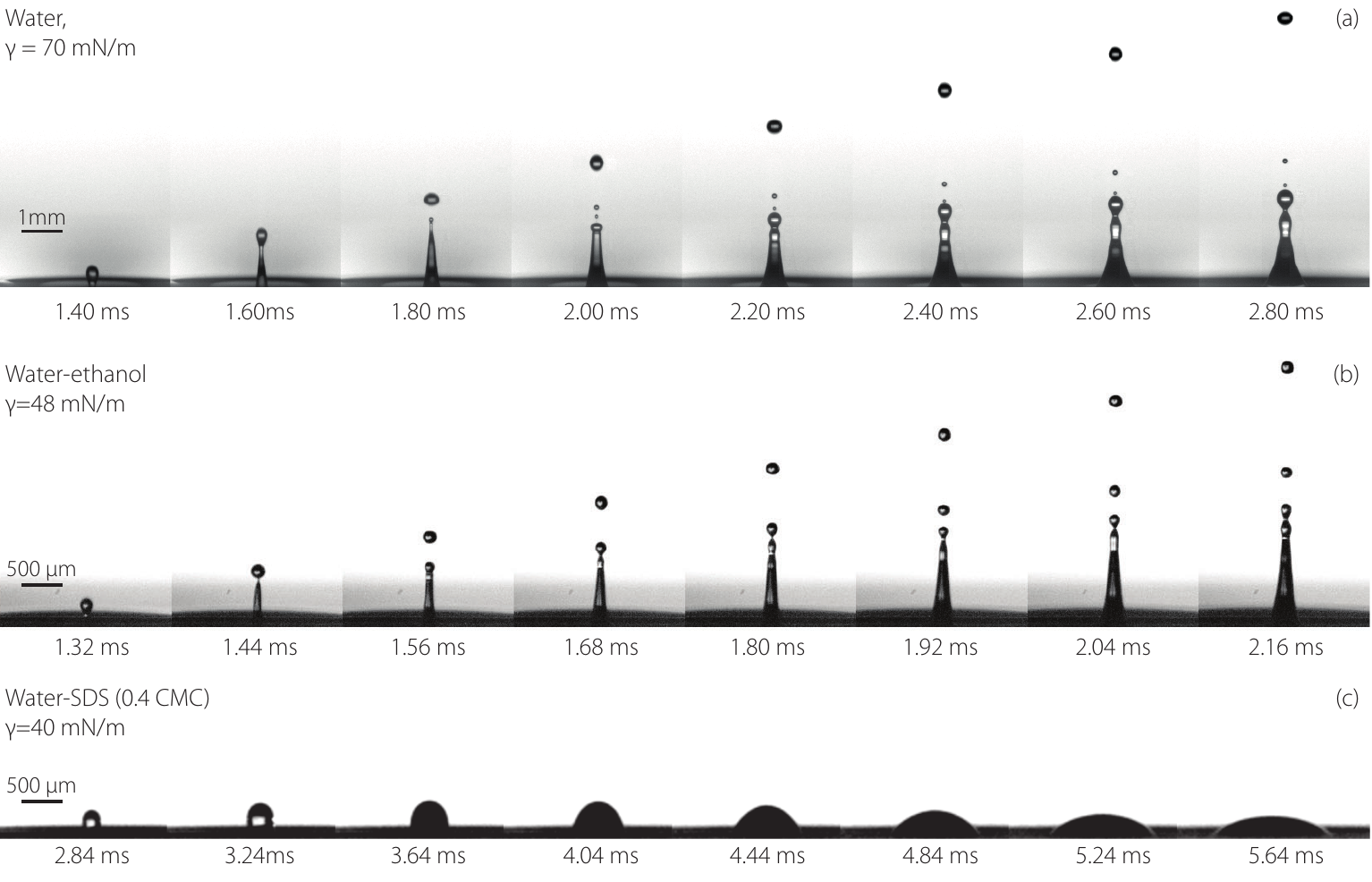}
    \caption{Sequences of bursting bubbles (bubbles start to burst at 0~ms) with comparable radius in three different liquids : (a) water, $\gamma$ = 70 mN.m$^{-1}$, R = 830 $\mu$m ; (b) water-ethanol solution with respectively 89.5\% and 10.5\% of total weight, $\gamma$ = 48 mN.m$^{-1}$, R = 830 $\mu$m ; (c) water-SDS solution with 3.4 mM of SDS, $\gamma$ = 40 mN.m$^{-1}$, R = 840 $\mu$m. The surfactants have a strong influence on the jet dynamics, independently of their influence on the surface tension.}
    \label{Qualitatif}
\end{figure}

\section{Experimental setup}

The experiment consists in releasing a single air bubble from a submerged needle in a  liquid and recording the upward jet and released drops after the bubble bursts at the free surface. 
Air bubbles are generated in a parallelipedal glass tank (20~cm length, 14~cm width, 9.5~cm depth) filled with either tap water or an aqueous solution of SDS (Sodium Dodecyl sulfate - purchased from Sigma Aldrich) surfactant with a mass concentration ranging from 0.5~g/L to 10~g/L, i.e. 1.7\,mM to 34.7\,mM. For SDS at the ambient temperature the critical micelle concentration (CMC) is found to be around 8mM \cite{Thominet1987}, which means that the SDS concentration in our solutions varies from C= 0.2 CMC to 4.3 CMC.
Bubbles are generated using a syringe pump filled with air. Three different needles are used, with internal diameter varying from 0.08~mm to 1.5~mm enabling to create bubble with thee different radii : 0.8, 1.1, 1.7 $\pm$ 0.1~mm. The bubbles rise to the surface and briefly float before bursting.
Considering the elliptic shape of the floating bubble, we defined an equivalent radius as $R=(a^2b)^{1/3}$ with $a$ and $b$ respectively the semi-major and semi-minor axes of the ellipse. 
The surface tension of each solution is measured using the pendant drop technique~\cite{Berry2015}.

In all experiments a digital high speed camera ({\it Phantom} V2511) is used to image the rising jet and releasing drops from the side, above the free surface. 
In a few experiments a second digital high speed camera ({\it Photron} SA-5) is added to image the collapse of the submerged cavity below the free surface.


\section{Qualitative description}

Figure~\ref{Qualitatif} presents three sequences of bubble busting. In each case the bubble radius is almost constant and the liquid is different : (a) tap water ($\gamma$ = 70 mN.m$^{-1}$), (b) a water-ethanol mixture with a surface tension $\gamma$ = 48 mN.m$^{-1}$, and (c) a water-SDS solution with a surface tension $\gamma$ = 40 mN.m$^{-1}$. 
We observe on (a) and (b) that the decrease of surface tension does not affect much the drop size and velocity. These observations have been reported quantitatively in the literature \cite{Duchemin2002, Ghabache2014, Seon2017, Brasz2018a, Berny2020}. In the sequence (c), the bubble bursts in a liquid with surfactants concentrated at 0.4~CMC and with a surface tension very close to that of sequence (b). The result is remarkable. The presence of surfactants completely changes the jet dynamics, the jet velocity is so low that it can barely reach the free surface and cannot produce any droplet. In the following, our goal is to examine more quantitatively the influence of the surfactant concentration on the drop dynamics, and to look at where and how, in the cavity collapse process, the surfactants can have such a strong influence.

\section{The bubble collapse}

\begin{figure}[h!]
    \centering
    \includegraphics[width=0.8\linewidth]{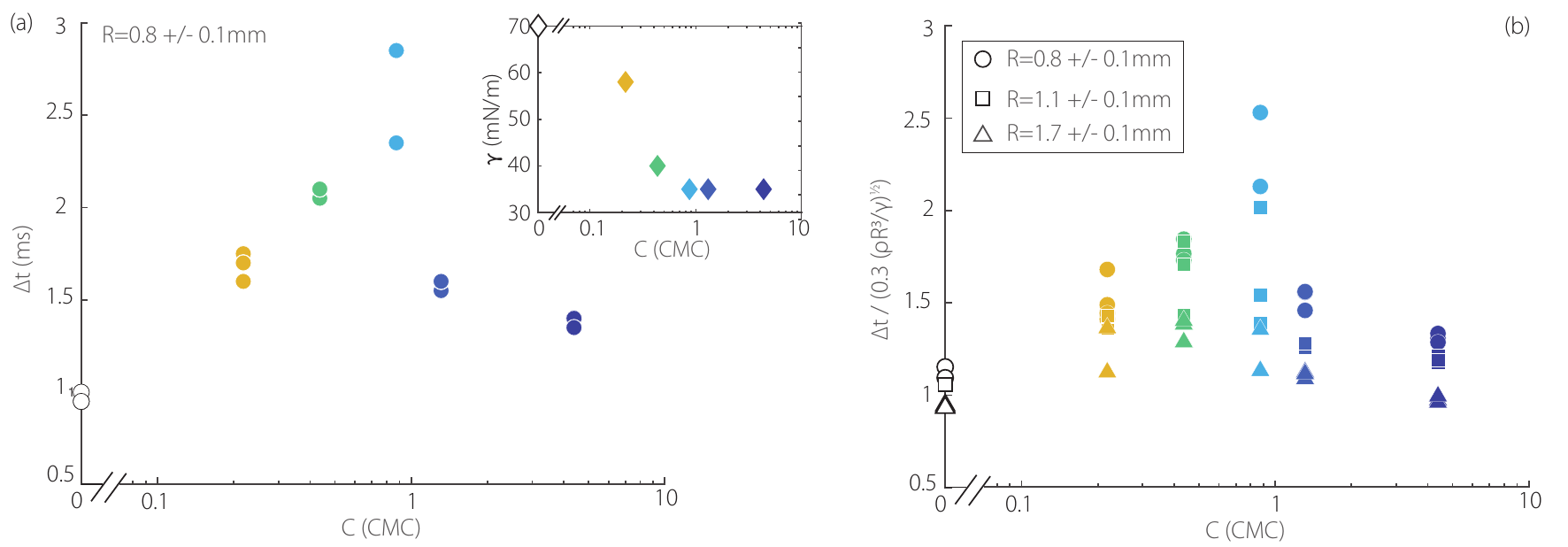}
    \caption{ (a) Bubble collapsing time $\Delta t$ as a function of the surfactant concentration, adimensionalized using the critical micelle concentration (CMC). The CMC is taken equal to 8~mM \cite{Thominet1987}. Inset: for each solution the surface tension is measured using the pendant drop technique and reported as a function of the surfactant concentration. (b) Collapsing time, $\Delta t$, normalized by the capillaro-inertial time scale, $\rho R^3/\gamma$, as a function of the dimensionless surfactant concentration, for three different bubble radius R : 0.8, 1.1 and 1.7 mm. The 0.3 prefactor is added in the normalization of $\Delta t$ in order to have the dimensionless collapsing time equal to one. The color of the markers is associated to the surfactant concentration in the liquid. }
    \label{CollapseTime}
\end{figure}

Before the cavity collapses, the bubble is floating at the free surface. As it is static, its shape is due to an equilibrium between capillarity and gravity and is obtained by integration of the Young–Laplace equation \cite{Toba1959, Lhuissier2011b, Poujol2021}. 
Surfactants have no more influence on the static bubble shape than through their modification of the surface tension. The static bubble shape before bursting cannot be responsible for the modification of the jet dynamics between figure~\ref{Qualitatif}(b) and (c). 

We then focus on the influence of the surfactant concentration on the bubble collapse duration $\Delta t$. $\Delta t$ is defined as
the time elapsed between the hole nucleation in the cap film and the cavity reversal, when the depth of the immersed cavity starts to decrease.
Figure~\ref{CollapseTime} (a) presents $\Delta t$ as a function of C(CMC) the surfactant concentration adimensionalized using the CMC. Note that the color of the markers is associated to the surfactant concentration in the liquid. The different markers at one concentration show $\Delta t$ for experiments in similar conditions, they therefore only reflect the dispersion of the results. We observe that, independently of the dispersion, the bubble collapsing time increases with the surfactant concentration, reaches a maximum close to the CMC and decreases. In other words, up to the CMC, the cavity is slower to collapse as the surfactants are more concentrated and, above the CMC, it becomes faster again. This non-monotonic variation of the collapsing time with the surfactant concentration is surprising. In particular because the variation of the surface tension $\gamma$ with the dimensionless surfactant concentration $C$ is monotonic, as verified in the inset that presents the variation of $\gamma$ with $C$. In this inset $\gamma$ expectedly decreases with $C$, from the water surface tension, until it reaches a plateau, around 36~mN/m, beyond the CMC. Consequently, the non-monotonic variation of $\Delta t$ with $C$ indicates that the surfactant dynamics should play a role in the cavity collapse.

In this capillaro-inertial collapse, $\Delta t$ is expected to scale as the capillaro-inertial time : $\rho R^3/\gamma$, with $\rho$ the liquid density and $R$ the bubble radius \cite{Poujol2021}.  Figure~\ref{CollapseTime} (b) presents the collapsing time $\Delta t$ normalized by this capillaro-inertial time scale as a function of the dimensionless surfactant concentration $C$, for the three different bubble radii $R =$ 0.8, 1.1 and 1.7~mm.  
As expected, without surfactant ($C=0$), all the bubbles with different radius collapse, demonstrating that this adimensionalized collapsing time is relevant. The prefactor 0.3 is added to the capillaro-inertial time so that the normalized times collapse around 1.
However, as the surfactants are added in solution, the dimensionless collapsing time increases and become more dispersed with increasing bubble size.
The dimensionless time and its dispersion both reach a maximum between half the CMC and the CMC (C= 0.5 - 1) and then decrease. As we approach the CMC, the experiments with different bubble radii do not collapse anymore and the small bubbles are relatively slower to collapse compared to the larger ones.

By normalizing $\Delta t$ by the relevant capillaro-inertial collapsing time, figure~\ref{CollapseTime} (b) enables to decorrelate the respective influence of the surface tension and the surfactant dynamics. As the data do not rescale in the presence of the surfactants, they are expected to be a consequence of the particular dynamics of the surfactants, independently of their influence on the measured liquid surface tension $\gamma$ displayed in the inset. Indeed, in processes that involve surface stretching and/or capillary waves, as it is the case in our experiment, gradients of surfactants can appear, generating Marangoni stresses that affect the dynamics \cite{Kamat2018, Manikantan2020}. Constante-Amores {\it et al.} \cite{Constante-Amores2021} showed that in the insoluble surfactant limit, the collapse yields to an over-concentration of surfactants at the apex of the cavity when the capillary waves focus, source of a strong Marangoni stress that can delay the cavity collapse. The presence of insoluble surfactants is also known to influence the dispersion of surface waves and to enhance capillary wave damping due to the interfacial rigidification \cite{Lucassen1966, Asaki1995}.

In the aim of contextualizing our results within the existing literature, we need to discuss the surfactant dynamics in our experiment. Two kinetics must be considered: the surface diffusivity and the surfactant rearrangement between the surface and the bulk.
First, the surface diffusivity $D_s$ of SDS is around 10$^{-9}$ m$^2$.s$^{-1}$ \cite{McGough2006}. Thus, the Peclet number $\text{Pe} = \sqrt{\gamma R /\rho}/D_s$ that measures the relative importance of surface convection of surfactant to its diffusion is $\mathcal{O}(10^4)$ in our experiment. The surfactant surface diffusion is therefore not strong enough to mitigate its advection.

Second, to estimate the adsorption-desorption dynamics of surfactants one can compare the characteristics time of the sorption rates to the diffusion time of the surfactants from the liquid-gas interface to the bulk.
The Langmuir model gives the characteristics time for the sorption kinetics : $\tau_b= \Gamma_m (k_\text{des}\Gamma_m+k_\text{ads}C)$ with $\Gamma_m = 4.10^{-6}$ mol.m$^{-2}$ \cite{Lu1995} the maximum surface packing concentration of SDS at the air-water interface and $k_\text{des}$ and $k_\text{ads}$ respectively the desorption and adsorption rate \cite{Chang1995}. 
The typical diffusive time-scale can be express as $t_\text{diff} =  \Gamma^2/(D_v C^2)$ with $\Gamma$ the  surface concentration at equilibrium \cite{Cantat2013} and $D_v = 5.10^{-10}$ m$^2$.s$^{-1}$ \cite{Kinoshita2017} the diffusion coefficient of the surfactants in the bulk liquid. 
Even if the characteristic time $\tau_b$ is not easy to estimate, in particular due to the lack of precision in the sorption rates $k_\text{des}$ and $k_\text{ads}$, the diffusion time, of the order of the millisecond, seems to remain larger than the time of the adsorption-desorption kinetics. This indicates that the dynamics of surfactants is limited by the diffusion, which is of the same order of magnitude than the time of collapse $\Delta t$. 
Thus, we expect to have a comparable time between surfactant dynamics and bubble collapse and, subsequently, to have solubility effects during the collapse. 
However, as the surprising variation of the collapsing time $\Delta t$ with the surfactant concentration $C$ is interpreted as a consequence of the variation of the local concentration of the surfactants, we expect the solubility not to be dominant. 
Consequently, the characteristics times of the surfactant dynamics might be underestimated, or the dynamics might be influenced by the presence of dodecanol, an insoluble impurity which may be either present from the original preparation or produced by hydrolysis of the aqueous SDS on standing \cite{Lu1995}. The use of numerical simulation would now be an asset in the aim of interpreting this complex situation \cite{Kamat2018,Constante-Amores2021}.

\section{Jet drops characteristics}

\begin{figure}[h!]
    \centering
    \includegraphics[width=\linewidth]{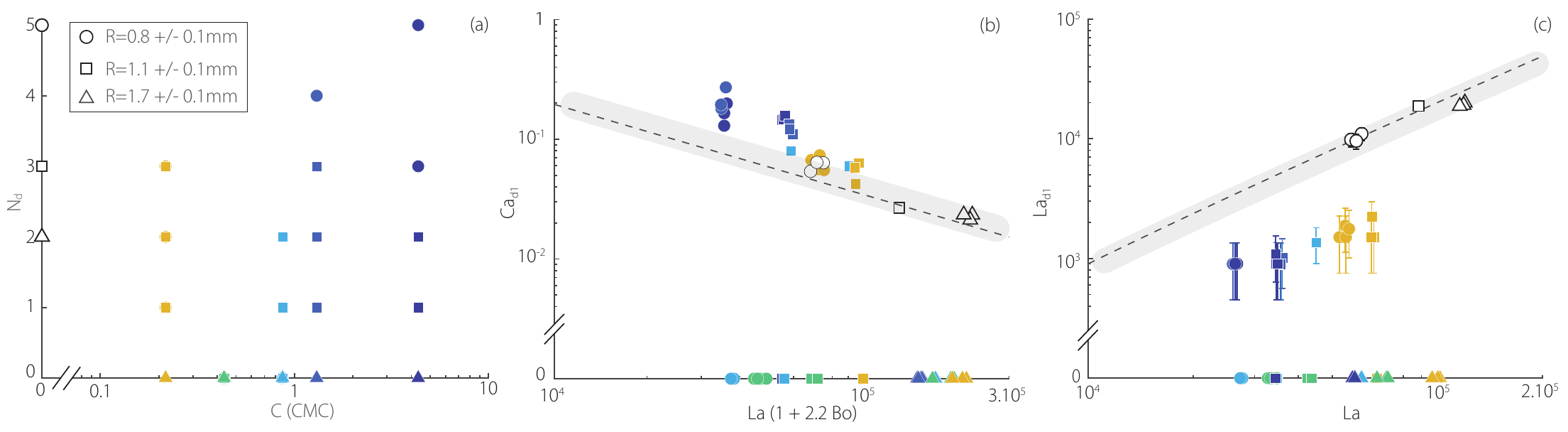}
    \caption{(a) Number of ejected drops as a function of the SDS concentration. (b) Capillary number of the first ejected drop as a function of La(1+2.2Bo) (c) Laplace number of the first ejected drop as function of the Laplace number of initial bubble. In all graphs color codes for the SDS concentration and symbol codes for the bubble size. In (b) and (c) the dashed lines represent the unified relations from~\cite{Deike2018}. The gray shadows traduce the dispersion of the experimental and numerical values all reported in \cite{Berny2020}.}
    \label{DropCharac}
\end{figure}

As shown in figure~\ref{Qualitatif}, after a bubble has collapsed, an upward jet usually rises and produces the so-called jet drops. Figure~\ref{DropCharac} (a) presents the number of these drops produced when a bubble bursts, as a function of the surfactant concentration, for 3 different bubble radii.  For bubbles bursting in water (C=0~CMC, empty markers) the 3 different bubbles produce droplets and the smaller the bubble, the more drops are produced \cite{Berny2020}. 
When surfactants are added to the water, our largest bubble size (triangle) cannot produce drops anymore, regardless of the surfactant concentration tested. This is a strong result, for this bubble size, surface contamination completely kills the drop production.  
Then, for smaller bubble, we observe that there is dispersion in the number of produced droplets. Indeed, for C $\simeq$ 0.2~CMC (yellow markers), R = 0.8 and 1.1~mm are superimposed and  produce either 1, 2 ou 3 droplets. There is the same kind of dispersion for C $\geq$ 0.9~CMC.
Nevertheless, despite this dispersion, the trend is clear:  \textit{(i)} surface contamination can prevent the drop production and \textit{(ii)} when droplets are produced there are less numerous and their number seems to decrease down to a minimum around half the CMC, before increasing again.  
These results are crucial, they signify that the size distribution of ejected jet drops produced in pure water \cite{Berny2021} might be very different than the one produced in water with surfactant. They precise and experimentally validate the recent numerical results of Constante-Amores \textit{et al.} \cite{Constante-Amores2021} that show that a reduction in the number of ejected droplets arises with surfactant-laden flow due to Marangoni flow.

To go further in influence of the surfactant contamination in the jet drop production, the speed and size of the first ejected droplet are quantified as a function of the surfactant concentration. Based on a large amount of numerical and experimental results, previous studies have demonstrated that the problem has two control parameters : the main one, the Laplace number (La), which compares the capillaro-inertial forces with the viscous forces, and the Bond number (Bo), which compares the gravitational forces with the capillary ones \cite{Duchemin2002, Ganan-Calvo2017, Deike2018, Berny2020}. They are defined as: 
\begin{align}
\text{La} & =  \frac{\rho \gamma R}{\mu^2} \\
\text{Bo} & =  \frac{\rho g R^2}{\gamma}
\end{align}
where $R$ is the bubble radius, $\mu$ the water viscosity, $\rho$ the water density, $\gamma$ the surface tension and $g$ the acceleration of gravity. The first drop speed $V_d$ and size $R_d$ are also adimensionalized using, respectively, the visco-capillary velocity $V_\mu=\mu/\gamma$ and length $l_\mu = \mu^2/(\rho \gamma)$, yielding the dimensionless drop speed and size:
\begin{align}
\text{Ca}_d & = \frac{V_d \mu}{\gamma}  \\
\text{La}_d & = \frac{\rho\gamma R_d}{\mu^2} 
\end{align}
Within this dimensionless framework, previous studies~\cite{Deike2018, Berny2020} have proposed universal rescalings able to fully describe the first drop velocity and size. These scalings are respectively represented with dashed line on figures~\ref{DropCharac} (b) and (c). They gather a large range of bubble size and liquid parameters ($\rho$, $\mu$, $\gamma$). The grey zone around the dashed line represents the error bar including all the experiences.
The Bond number appears in the x-axis as a correction term for the drop velocity (Ca$_d$) and plays no role for the drop size (La$_d$). 

On these plots we add here the values measured with bubbles bursting in our solutions of SDS mixed to water. The surfactant concentration is represented using the same colors as in figures~\ref{CollapseTime} and \ref{DropCharac}(a).
First, expectedly, the drop velocity and size from bubble busting in water, with empty markers, fall onto the universal scalings, and many markers are on the x-axis because no drop are produced for the largest bubble and for the concentration close to half of the CMC, as shown in figure~\ref{DropCharac}(a).
Secondly, in figure~\ref{DropCharac}(b), we observe that the more concentrated is the solution, the higher the drop velocity is above the scaling.
In figure~\ref{DropCharac}(c), even with a small amount of surfactant, the drop size falls quite far below the universal scaling. It also seems that the drop size is less affected by the Laplace number as the surfactant concentration is high. 
These variations indicate that the influence of the surfactants is highly non trivial, undoubtedly dependent on the local gradient concentration along the jet.
Finally, we observe again that the dispersion is quite larger than without surfactant, may be because the dynamics is very sensitive to the balance between the coupled dynamics of surfactants and the jet. Consequently, the next step of this study will probably need a statistical characterization to properly capture the influence of the surfactant concentration in the drop production \cite{Berny2021a}.

\section{Capillary waves focusing}

\begin{figure}[h!]
    \centering
    \includegraphics[width=1\linewidth]{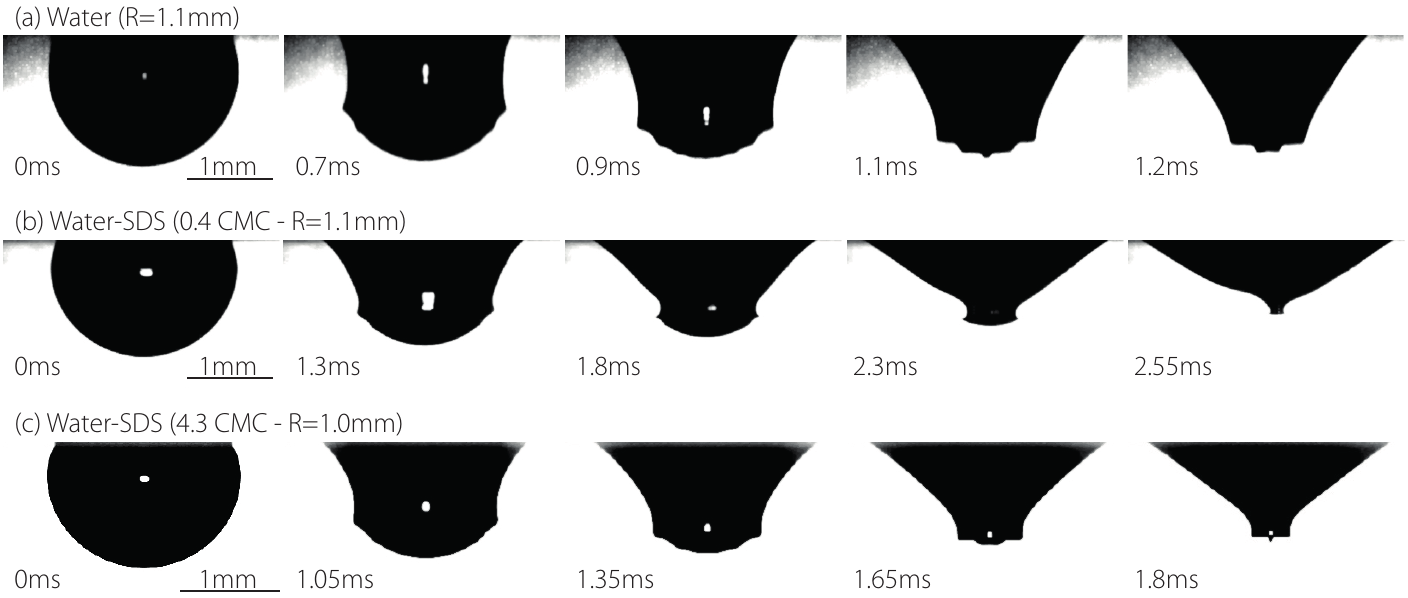}
    \caption{Sequences showing the collapse of the submerged cavity for bubbles with almost the same radius in three different liquids: (a) water, $\gamma=70\,$mN.m$^{-1}$, $R=1.1\,$mm, (b) water-SDS solution at 0.4\,CMC, $R=1.1\,$mm, (c) water-SDS solution at 4.3\,CMC, $R=1.0\,$mm.}
    \label{SeqCollapse}
\end{figure}

The jet dynamics strongly depends on the capillary waves focusing at the bottom of the cavity \cite{Ghabache2014, Gordillo2019}. 
Figure~\ref{SeqCollapse} presents 3 sequences of cavity collapse with almost the same bubble radius and 3 different surfactant concentrations: (a) no surfactant, (b) 0.4~CMC and (c) 4.3~CMC. The capillary waves propagation is different in these 3 sequences. 
The clearest difference appears between (a) and (b), and lies in the wave shape, in particular in the second half of the collapse sequences. The shape of the lowest collapsing cavity shown in the second to last images is completely different, and undoubtedly explains the strong difference in the drop production dynamics (see $C=0.4$~CMC in figure~\ref{DropCharac}). 
As shown by Constante-Amores \textit{et al.} \cite{Constante-Amores2021} for precise values of the control parameters, the interfacial surfactant concentration reaches its maximum value as the surfactant-laden capillary waves converge on the cavity apex. The Marangoni stresses that drive motion from high to low surface concentration regions can explain the shape of the cavity. 

For the highest concentration  (c) the shape of the capillary waves, and of the lowest cavity, looks quite similar to the one in water. 
This seems to indicate that for concentration higher than the CMC, the Marangoni stresses are lower, due to lower concentration gradient, and this can be explained by a smaller diffusive time of the surfactants for high concentration, giving a surfactant soluble behavior to the collapse. 
This similarity in the cavity collapse between no contamination and high contamination probably explains why the drop characteristics are closer between water and the highest concentration than water and 0.4~CMC, for which there is no drop produced.

\section{Conclusion}

Surfactants are most often present in the liquids where bubbles are bursting (ocean, sparkle wine, soda...) and have barely been taken into account in the experiments and the models. In particular, to our knowledge, experiments of a single bubble bursting in a surfactant laden liquid have never been carried out. Here, we have shown how the SDS strongly influences the cavity collapse and the drops production, for different values of the bubble size and of the surfactant concentration. In particular, we highlight that the contamination induces: (i) a maximum in the bubble collapse duration around the CMC, (ii) smaller and faster drops and (iii) less drops, with no drop at all for a particular concentration of half the CMC. We also show that these effects are a consequence of the surface tension gradients (Marangoni stresses) and not just the surface tension lowering. The exact role of the Marangoni flows is not known and needs to be clarified by quantifying the surface tension gradients appearing during the bubble collapse. 

In the following, motivated by this study, more experiments should be done, in particular with insoluble surfactants to examine the influence of the solubility. As it is complex to only change one parameter at a time, it would undoubtedly be interesting to carry out a large campaign of numerical simulations. Indeed, the cavity collapse, cavity reversal, jet dynamics and end pinching are very complex phenomenon and their dynamics involve a high dependency on the local gradient concentration. In simulations Marangoni stresses can be turned off while surfactant-induced lowering of surface tension can be retained, thereby determining which of the two effects is the dominant mechanism by which surfactants affect the flow \cite{Kamat2018}. On the other hand, simulations would also enable a statistical characterization of the drop production \cite{Berny2021a} as it seems to be very sensitive to the experimental conditions and probably to the initial conditions.
Finally, in any way, the surfactant concentration needs to be taken into account in the experiments and simulations to improve the prediction of a real bubble bursting spray as the sea spray.

\bibliography{BubbleSurfactant}

\begin{thebibliography}{10}

\bibitem{Garrett2000}
Chris Garrett, Ming Li, and David Farmer.
\newblock The connection between bubble size spectra and energy dissipation
  rates in the upper ocean.
\newblock {\em Journal of physical oceanography}, 30(9):2163--2171, 2000.

\bibitem{Deane2002}
Grant~B Deane and M~Dale Stokes.
\newblock Scale dependence of bubble creation mechanisms in breaking waves.
\newblock {\em Nature}, 418(6900):839--844, 2002.

\bibitem{Prather2013}
Kimberly~A Prather, Timothy~H Bertram, Vicki~H Grassian, Grant~B Deane, M~Dale
  Stokes, Paul~J DeMott, Lihini~I Aluwihare, Brian~P Palenik, Farooq Azam,
  John~H Seinfeld, et~al.
\newblock Bringing the ocean into the laboratory to probe the chemical
  complexity of sea spray aerosol.
\newblock {\em Proceedings of the National Academy of Sciences},
  110(19):7550--7555, 2013.

\bibitem{Deike2016}
Luc Deike, W~Kendall Melville, and St{\'e}phane Popinet.
\newblock Air entrainment and bubble statistics in breaking waves.
\newblock {\em Journal of Fluid Mechanics}, 801:91--129, 2016.

\bibitem{Deike2021}
Luc Deike.
\newblock Mass transfer at the ocean--atmosphere interface: The role of wave
  breaking, droplets, and bubbles.
\newblock {\em Annual Review of Fluid Mechanics}, 54, 2021.

\bibitem{Blanchard1988}
D.~C. Blanchard and L.~D. Syzdek.
\newblock Film drop production as a function of bubble size.
\newblock {\em J. Geophys. Res.}, 93(C4):3649--3654, 1988.

\bibitem{Lhuissier2011b}
H.~Lhuissier and E.~Villermaux.
\newblock Bursting bubble aerosols.
\newblock {\em Journal of Fluid Mechanics}, 696:5--44, 2011.

\bibitem{Spiel1997}
D.~E. Spiel.
\newblock More on the births of jet drops from bubbles bursting on seawater
  surfaces.
\newblock {\em J. Geophys. Res.}, 102(C3):5815--5821, 1997.

\bibitem{Ghabache2014}
Elisabeth Ghabache, Arnaud Antkowiak, Christophe Josserand, and Thomas
  S{\'e}on.
\newblock On the physics of fizziness: How bubble bursting controls droplets
  ejection.
\newblock {\em Physics of Fluids (1994-present)}, 26(12):--, 2014.

\bibitem{Ghabache2016a}
E.~Ghabache and T.~S{\'e}on.
\newblock Size of the top jet drop produced by bubble bursting.
\newblock {\em Phys. Rev. Fluids}, 1(051901), 2016.

\bibitem{Brasz2018a}
C~Frederik Brasz, Casey~T Bartlett, Peter~LL Walls, Elena~G Flynn,
  Yingxian~Estella Yu, and James~C Bird.
\newblock Minimum size for the top jet drop from a bursting bubble.
\newblock {\em Physical Review Fluids}, 3(7):074001, 2018.

\bibitem{Ganan-Calvo2017}
Alfonso~M Ga{\~n}{\'a}n-Calvo.
\newblock Revision of bubble bursting: universal scaling laws of top jet drop
  size and speed.
\newblock {\em Physical review letters}, 119(20):204502, 2017.

\bibitem{Blanco--Rodriguez2020}
Francisco~J Blanco-Rodr{\'\i}guez and JM~Gordillo.
\newblock On the sea spray aerosol originated from bubble bursting jets.
\newblock {\em Journal of Fluid Mechanics}, 886, 2020.

\bibitem{Veron2012}
F~Veron, C~Hopkins, EL~Harrison, and JA~Mueller.
\newblock Sea spray spume droplet production in high wind speeds.
\newblock {\em Geophysical Research Letters}, 39(16), 2012.

\bibitem{Lewis2004}
E.~R. Lewis and S.~E. Schwartz.
\newblock {\em Sea Salt Aerosol Production. Mechanisms, Methods, Measurements,
  and Models}.
\newblock American Geophysical Union, Washington, DC,, geophysical monograph
  152. edition, 2004.

\bibitem{Veron2015}
Fabrice Veron.
\newblock Ocean spray.
\newblock {\em Annual Review of Fluid Mechanics}, 47(1):507--538, 2015.

\bibitem{Leeuw2011}
Gerrit de~Leeuw, Edgar~L Andreas, Magdalena~D. Anguelova, C.~W. Fairall,
  Ernie~R. Lewis, Colin O'Dowd, Michael Schulz, and Stephen~E. Schwartz.
\newblock Production flux of sea spray aerosol.
\newblock {\em Rev. Geophys.}, 49(2), 05 2011.

\bibitem{Deike2018a}
Luc Deike and W~Kendall Melville.
\newblock Gas transfer by breaking waves.
\newblock {\em Geophysical Research Letters}, 45(19):10--482, 2018.

\bibitem{Blanchard1963}
D.~C Blanchard.
\newblock The electrification of the atmosphere by particles from bubbles in
  the sea.
\newblock {\em Progress In Oceanography}, 1:73 -- 112, IN7, 113--202, 1963.

\bibitem{Seon2017}
Thomas S{\'e}on and G{\'e}rard Liger-Belair.
\newblock Effervescence in champagne and sparkling wines: From bubble bursting
  to droplet evaporation.
\newblock {\em The European Physical Journal Special Topics}, 226(1):117--156,
  2017.

\bibitem{Duchemin2002}
L.~Duchemin, S.~Popinet, C.~Josserand, and S.~Zaleski.
\newblock Jet formation in bubbles bursting at a free surface.
\newblock {\em Phys. Fluids}, 14(9):3000--3008, 2002.

\bibitem{Berny2020}
Alexis Berny, Luc Deike, Thomas S{\'e}on, and St{\'e}phane Popinet.
\newblock Role of all jet drops in mass transfer from bursting bubbles.
\newblock {\em Physical Review Fluids}, 5(3):033605, 2020.

\bibitem{Ganan-Calvo2018}
Alfonso~M. Ganan-Calvo.
\newblock Scaling laws of top jet drop size and speed from bubble bursting
  including gravity and inviscid limit.
\newblock {\em Phys. Rev. Fluids}, 3:091601, Sep 2018.

\bibitem{Berny2021}
Alexis Berny, St{\'e}phane Popinet, Thomas S{\'e}on, and Luc Deike.
\newblock Statistics of jet drop production.
\newblock {\em Geophysical Research Letters}, 48(10), 2021.

\bibitem{Wurl2011}
Oliver Wurl, E~Wurl, L~Miller, K~Johnson, and Svein Vagle.
\newblock Formation and global distribution of sea-surface microlayers.
\newblock {\em Biogeosciences}, 8(1):121--135, 2011.

\bibitem{Poulain2018a}
S~Poulain, E~Villermaux, and L~Bourouiba.
\newblock Ageing and burst of surface bubbles.
\newblock {\em Journal of fluid mechanics}, 851:636--671, 2018.

\bibitem{Shaw2021}
Daniel~B Shaw and Luc Deike.
\newblock Surface bubble coalescence.
\newblock {\em Journal of Fluid Mechanics}, 915, 2021.

\bibitem{Neel2021}
B~N{\'e}el and L~Deike.
\newblock Collective bursting of free-surface bubbles, and the role of surface
  contamination.
\newblock {\em Journal of Fluid Mechanics}, 917, 2021.

\bibitem{Modini2013}
RL~Modini, LM~Russell, GB~Deane, and MD~Stokes.
\newblock Effect of soluble surfactant on bubble persistence and
  bubble-produced aerosol particles.
\newblock {\em Journal of Geophysical Research: Atmospheres},
  118(3):1388--1400, 2013.

\bibitem{Quinn2015}
Patricia~K Quinn, Douglas~B Collins, Vicki~H Grassian, Kimberly~A Prather, and
  Timothy~S Bates.
\newblock Chemistry and related properties of freshly emitted sea spray
  aerosol.
\newblock {\em Chemical reviews}, 115(10):4383--4399, 2015.

\bibitem{Constante-Amores2021}
Cristian~R Constante-Amores, Lyes Kahouadji, Assen Batchvarov, Seungwon Shin,
  Jalel Chergui, Damir Juric, and Omar~K Matar.
\newblock Dynamics of a surfactant-laden bubble bursting through an interface.
\newblock {\em Journal of Fluid Mechanics}, 911, 2021.

\bibitem{Thominet1987}
V.~Thominet, C.~Stenvot, and D.~Langevin.
\newblock Light scattering study of the viscoelasticity of soluble monolayers.
\newblock {\em Journal of colloid and Interface Science}, 126(1):54, 1987.

\bibitem{Berry2015}
Joseph~D. Berry, Michael~J. Neeson, Raymond~R. Dagastine, Derek~Y.C. Chan, and
  Rico~F. Tabor.
\newblock Measurement of surface and interfacial tension using pendant drop
  tensiometry.
\newblock {\em Journal of colloid and interface science}, 454:226--237, 2015.

\bibitem{Toba1959}
Yoshiaki Toba.
\newblock Drop production by bursting of air bubbles on the sea surface (ii)
  theoretical study on the shape of floating bubbles.
\newblock {\em Journal of the Oceanographical Society of Japan},
  15(3):121--130, 1959.

\bibitem{Poujol2021}
Mathis Poujol, R{\'e}gis Wunenburger, Fran{\c{c}}ois Ollivier, Arnaud
  Antkowiak, and Juliette Pierre.
\newblock Sound of effervescence.
\newblock {\em Physical Review Fluids}, 6(1):013604, 2021.

\bibitem{Kamat2018}
Pritish~M Kamat, Brayden~W Wagoner, Sumeet~S Thete, and Osman~A Basaran.
\newblock Role of marangoni stress during breakup of surfactant-covered liquid
  threads: reduced rates of thinning and microthread cascades.
\newblock {\em Physical Review Fluids}, 3(4):043602, 2018.

\bibitem{Manikantan2020}
Harishankar Manikantan and Todd~M. Squires.
\newblock Surfactant dynamics: hidden variables controlling fluid flows.
\newblock {\em Journal of Fluid Mechanics}, 892, apr 2020.

\bibitem{Lucassen1966}
J~Lucassen and Robert~S Hansen.
\newblock Damping of waves on monolayer-covered surfaces: I. systems with
  negligible surface dilational viscosity.
\newblock {\em Journal of Colloid and Interface Science}, 22(1):32--44, 1966.

\bibitem{Asaki1995}
TJ~Asaki, DB~Thiessen, and PL~Marston.
\newblock Effect of an insoluble surfactant on capillary oscillations of
  bubbles in water: observation of a maximum in the damping.
\newblock {\em Physical review letters}, 75(14):2686, 1995.

\bibitem{McGough2006}
Patrick~T McGough and Osman~A Basaran.
\newblock Repeated formation of fluid threads in breakup of a
  surfactant-covered jet.
\newblock {\em Physical review letters}, 96(5):054502, 2006.

\bibitem{Lu1995}
JR~Lu, IP~Purcell, EM~Lee, EA~Simister, RK~Thomas, AR~Rennie, and J~Penfold.
\newblock The composition and structure of sodium dodecyl sulfate-dodecanol
  mixtures adsorbed at the air-water interface: a neutron reflection study.
\newblock {\em Journal of Colloid and Interface science}, 174(2):441--455,
  1995.

\bibitem{Chang1995}
Chien-Hsiang Chang and Elias~I Franses.
\newblock Adsorption dynamics of surfactants at the air/water interface: a
  critical review of mathematical models, data, and mechanisms.
\newblock {\em Colloids and Surfaces A: Physicochemical and Engineering
  Aspects}, 100:1--45, 1995.

\bibitem{Cantat2013}
Isabelle Cantat, Sylvie Cohen-Addad, Florence Elias, Fran{\c{c}}ois Graner,
  Reinhard H{\"o}hler, Olivier Pitois, Florence Rouyer, and Arnaud
  Saint-Jalmes.
\newblock {\em Foams: structure and dynamics}.
\newblock OUP Oxford, 2013.

\bibitem{Kinoshita2017}
Koji Kinoshita, Elisa Parra, and David Needham.
\newblock Adsorption of ionic surfactants at microscopic air-water interfaces
  using the micropipette interfacial area-expansion method: Measurement of the
  diffusion coefficient and renormalization of the mean ionic activity for sds.
\newblock {\em Journal of colloid and interface science}, 504:765--779, 2017.

\bibitem{Deike2018}
Luc Deike, Elisabeth Ghabache, G{\'e}rard Liger-Belair, Arup~K Das,
  St{\'e}phane Zaleski, St{\'e}phane Popinet, and Thomas S{\'e}on.
\newblock Dynamics of jets produced by bursting bubbles.
\newblock {\em Physical Review Fluids}, 3(1):013603, 2018.

\bibitem{Berny2021a}
Alexis Berny, Luc Deike, St{\'e}phane Popinet, and Thomas S{\'e}on.
\newblock How size and speed of jet drops are robust to initial conditions.
\newblock {\em Subm. to Physical Review Fluids}, 2021.

\bibitem{Gordillo2019}
JM~Gordillo and J~Rodr{\'\i}guez-Rodr{\'\i}guez.
\newblock Capillary waves control the ejection of bubble bursting jets.
\newblock {\em Journal of Fluid Mechanics}, 867:556--571, 2019.

\end{thebibliography}

\end{document}